\documentclass[12pt]{article}
\usepackage{amsmath}
\usepackage{geometry}                
\usepackage{graphicx}
\usepackage{amssymb}
\usepackage{epstopdf}
\def\cala{\cal A}

\def\cala{{\cal A}}

\def\call{{\cal L}}
\def\calr{{\cal R}}
\def\calrr{{\cal R}^{(2)}}
\def\calrn{{\cal R}^{(n)}}
\def\calrk{{\cal R}^{(k)}}
\def\calgk{{\cal G}^{(k)}}
\def\calck{{\cal C}^{(k)}}

\def\calaa{{\cal A}^{(2)}}

\def\callk{{\cal L}^{(k)}}
\def\calg{{\cal G}}

\def\b2hat{ {\hat b}_2 }

\begin{document}

\begin{titlepage}
\vfill
\begin{flushright}
\end{flushright}

\vfill
\begin{center}
\baselineskip=16pt
{\Large\bf The Riemann-Lovelock Curvature Tensor}
\vskip 0.15in
\vskip 0.5cm
{\large {\sl }}
\vskip 10.mm
{\large\bf 
David Kastor} 

\vskip 0.5cm
{\bf{
	Department of Physics\\ University of Massachusetts\\ Amherst, MA 01003\\	
	\texttt{kastor@physics.umass.edu}
     }}
\vspace{6pt}
\end{center}
\vskip 0.2in
\par
\begin{center}
{\bf Abstract}
 \end{center}
\begin{quote}
In order to study the properties of Lovelock gravity theories in low dimensions, we define
the $k$th-order Riemann-Lovelock tensor as a certain quantity having a total $4k$-indices, which is $k$th order in the Riemann curvature tensor and shares its basic algebraic and differential properties.  We show that the $k$th order Riemann-Lovelock tensor is determined by its traces in dimensions $2k\le D <4k$.  In $D=2k+1$ this identity implies that all solutions of pure $k$th order Lovelock gravity are `Riemann-Lovelock' flat.  It is verified that the static, spherically symmetric solutions of these theories, which are missing solid angle space times, indeed satisfy this flatness property.  This generalizes results from Einstein gravity in $D=3$, which corresponds to the $k=1$ case.  We speculate about some possible further consequences of Riemann-Lovelock curvature.
  \vfill
\vskip 2.mm
\end{quote}
\end{titlepage}


\section{Introduction}

Lovelock gravity \cite{Lovelock:1971yv} is the most general extension of  Einstein gravity to higher dimensions with field equations that depend, as in general relativity,  only on the curvature tensor and not on its derivatives.  It follows from this that Lovelock gravity also shares other important physical properties with general relativity.  Namely, it has a reasonably well behaved initial value formulation \cite{lovelock-hamiltonian} and admits constant curvature vacua with ghost-free excitation spectra \cite{Boulware:1985wk}.  Although Lovelock theory does not include the full range of higher curvature interactions present in an effective field theory treatment, it has nonetheless enjoyed great interest (see {\it e.g.} the reviews \cite{Charmousis:2008kc,Garraffo:2008hu} on Lovelock black hole solutions).

Given that Lovelock gravity shares certain important properties with Einstein gravity, it is natural to ask how far this similarity goes.  In this paper, we will ask a question about similarities, or lack thereof, regarding how the physics of Lovelock gravity  low dimensions.   

It is well known that Einstein gravity simplifies to drastic, but differing, extents in all dimensions below four.  With a single dimension, there simply is no intrinsic curvature and gravity in the sense of general relativity does not exist.  In two dimensions, intrinsic curvature exists, characterized by a single independent component of the Riemann tensor, but the Einstein tensor vanishes identically.  Any metric, therefore,  satisfies the vacuum Einstein equations in $D=2$ and the theory is trivial.

Three dimensions provides the first interesting case for Einstein gravity.  As discussed in {\it e.g.} reference \cite{Witten:1988hc}, it is almost, but not quite, trivial.
The independent components of the Riemann tensor in $D=3$ number the same as those of its contraction, the Ricci tensor, and the Riemann tensor may therefore be expressed in terms of the Ricci tensor and scalar curvature.  The explicit expression is easily obtained by considering the quantity 
\begin{equation}\label{basic}
A_{ab}{}^{cd}\equiv \delta_{abgh}^{cdef}\,  R_{ef}{}^{gh},
\end{equation}
where the delta-symbol is totally antisymmetrized 
and of overall strength one. On one hand, the quantity $A_{ab}{}^{cd}$ vanishes in $D=3$ because it involves antisymmetrization over $4$ indices.  While, on the other hand, it is  given in any dimension by 
\begin{equation}\label{3identity}
A_{ab}{}^{cd} = {1\over 6}\left(  R_{ab}{}^{cd} - 4 \delta^{[c}_{[a}R_{b]}{}^{d]} +  \delta^{cd}_{ab} \, R\right).
\end{equation}
It follows from this reasoning that in $D=3$ the Riemann tensor may be expressed as
\begin{equation}\label{curv}
R_{ab}{}^{cd} = 4 \delta^{[c}_{[a}R_{b]}{}^{d]} - \delta^{cd}_{ab} \, R.
\end{equation}
In vacuum gravity, the field equations imply that $R_{ab}$ and $R$ vanish,  and therefore in $D=3$ the vacuum field equations imply that the Riemann tensor vanishes as well.  All solutions of vacuum gravity in $D=3$ are flat.  In particular, this implies that there is no gravitational radiation\footnote{Consistent with this, the identity (\ref{curv}) also implies that the Weyl tensor, which signals the presence of gravity waves, vanishes identically in $D=3$}, or stated in another way, that three dimensional gravity has no local degrees of freedom.  

There are, however, important global effects.  Point masses in $D=3$ are represented by conical space times that are locally, but not globaly flat \cite{Deser:1983tn}.  The mass is proportional to the non-trivial holonomy of vectors parallel transported around the cone.  
There are also, famously, black holes if one includes a negative cosmological constant \cite{Banados:1992wn}.

Lovelock analogues of how Einstein gravity behaves in $D=1$ and $D=2$ are well known.
The Lovelock Lagrangian is a sum of higher curvature terms, which can be written as
\begin{equation}\label{lovelock}
\call = \sum_{k=0}^\infty c_k\call^{(k)},
\end{equation}
Here each $\call^{(k)}$ is a scalar combination of $k$ powers of the Riemann tensor that has a certain `quasi-topological' character, and the $c_k$'s are a set of coupling constants.  Specifically, the integral of $\call^{(k)}$ over a compact manifold of dimension $D=2k$ gives the manifold's Euler character, a topological quantity that is independent of the choice of metric.  For example, the $k=1$ term in (\ref{lovelock}) is simply the Einstein Lagrangian $\call^{(1)}=R$, which integrates to give the Euler character in $D=2$.  

Since its integral is independent of the metric, the $k$th Lovelock interaction makes no contribution to the field equations in $D=2k$.  The Einstein-like tensor $\calg^{(k)}_{ab}$ that follows from varying $\sqrt{-g}\call^{(k)}$ with respect to the metric in any dimension vanishes identically in $D=2k$.  This  provides the higher curvature generalization of the behavior of the Einstein tensor in $D=2$.  Meanwhile, the analogue of the triviality of Einstein gravity in $D=1$ stems from the fact that the interaction $\call^{(k)}$ itself vanishes for $D<2k$.  Like intrinsic curvature in $D=1$, there simply is no $k$th order Lovelock interaction in these lower dimensions.

It remains to say whether there is a Lovelock parallel of the, almost but not quite trivial, behavior of Einstein gravity in $D=3$ which stems from the curvature identity (\ref{curv}).  So far as we know, no such higher curvature analogue of this identity has appeared in the literature.  A simple reason is the absence within the Lovelock gravity formalism  of an analogue for the Riemann tensor.  As noted above,  an Einstein-like tensor $\calg^{(k)}_{ab}$ (that is  $k$th order in the actual curvature) is obtained by varying the $k$th order term in the Lovelock Lagrangian.  This tensor shares the hallmarks of the actual Einstein tensor - it is symmetric, satisfies a Bianchi identity $\nabla_b \calg^{(k)}_{a}{}^b=0$, and its trace is proportional to the corresponding term $\call^{(k)}$ in the Lagrangian.  Moreover, the Einstein-like tensor can be separated into contributions $\calrk_a{}^b$ and $\calrk$ from $k$th order analogues  of the Ricci tensor and scalar curvature such that $\calrk_a{}^a=\calrk$.  However, it is not known whether the Ricci-like tensor $\calrk_a{}^b$ can be expressed as the  trace of any quantity that has properties analogous to those of the Riemann tensor\footnote{See, however, the important note added regarding references \cite{Dadhich:2008df,Dadhich:2012cv} following the body of the paper.}.

In this paper, we will find such a higher curvature analogue of the Riemann tensor, which will be referred to as the Riemann-Lovelock tensor.  The $k$th order Riemann-Lovelock tensor, which is $k$th order in the underlying Riemann tensor,  has a total of $4k$ indices, has symmetries analogous to those of the actual Riemann tensor and satisfies a Bianchi identity.  We then study the curvature identities satisfied in low dimensions by the Riemann-Lovelock tensor and its traces.  These are analogues of  equation (\ref{curv}), which will, at least in part, answer the question regarding the extent to which Lovelock gravity has properties analogous to that generalize those of Einstein gravity in $D=3$.  

In a sense we find more than we were looking for.  The $k$th order Riemann tensor is determined by its traces in all dimensions $2k\le D<4k$.   In $D=2k+1$ it is determined by the Ricci-like tensor $\calrk_a{}^b$.
We will comment on the implications of this for solutions of Lovelock gravity in dimensions $D=2k+1$.  The implications are strongest for `pure' $k$th order Lovelock gravity, with Lagrangian $\call=\callk$.  In this case, the static, spherically symmetric solutions in $D=2k+1$ are missing solid angle space times.  These have non-trivial Riemannian curvature, but are `Riemann-Lovelock flat'  in the sense of having vanishing Riemann-Lovelock tensor.  We conclude by noting some further possible consequences of the notion of Riemann-Lovelock curvature, which are left as areas for future investigation.

\section{Lovelock formulas}\label{lovelocksection}

At this point, we need to introduce the formulas for the terms in the Lovelock Lagrangian (\ref{lovelock}).  Let us start by defining the quantity
\begin{equation}\label{lovelockterms}
\calrk = 
\delta^{a_1\dots a_{2k}}_{b_1\dots b_{2k}} \, 
R_{a_1a_2} {}^{b_1b_2}\,\dots\,  R_{a_{2k-1}a_{2k}} {}^{b_{2k-1}b_{2k}},
\end{equation}
where 
the delta-symbol is totally anti-symmetric and normalized to have unit strength, so that 
$\delta_{a_1\dots a_n}^{b_1\dots b_n}=\delta_{[a_1}^{b_1}\cdots\delta_{a_n]}^{b_n}$.
It follows immediately that $\calrk=0$ for $D<2k$ because of the number of indices antisymmetrized over.
The $k$th order term in the Lovelock Lagrangian is proportional to $\calrk$, with the conventional normalization being $\callk=\alpha_k\calrk$.
with $\alpha_k= (2k)!/2^k$.

Varying the individual terms in the Lovelock action 
yields the Einstein-like tensors
\begin{equation}\label{lovelockmotion}
\mathcal{G}^{(k)}_a{}^b = 
{(2k+1)\alpha_k\over 2}\, 
\delta^{bc_1\dots c_{2k}}_{ad_1\dots d_{2k}}\,R_{c_1c_2}{}^{d_1d_2}\, \dots\,R_{c_{2k-1}c_{2k}}{}^{d_{2k-1}d_{2k}},
\end{equation}
One can check that the tensors $\calg^{(k)}_{ab}$ are symmetric tensors and that the trace $\mathcal{G}^{(k)}_a{}^a$  is proportional to $\calrk$, and hence also to $\callk$.
It is clear from the antisymmetrization in (\ref{lovelockmotion})  that, in addition to vanishing for $D<2k$ where the $k$th order Lovelock term itself vanishes, the tensor 
$\mathcal{G}^{(k)}_a{}^b$ also vanishes identically in $D=2k$.  This is the higher curvature Lovelock analogue, noted above, of  the Einstein tensor vanishing in $D=2$.  The Bianchi identity $\nabla_a\calg^{(k)a}{}_{b}=0$ also follows from antisymmetry  together with the Bianchi identity for the underlying Riemann tensor.  The full equation of motion for Lovelock gravity is given by $\sum_{k=0}^\infty c_k\calg^{(k)a}{}_b=0$.

The $k$th order Einstein-like tensor (\ref{lovelockmotion}) may be written in a form, similar to the actual Einstein tensor, involving a Ricci-like tensor and a contribution proportional to the scalar quantity (\ref{lovelockterms}).  Writing the $\delta$ symbol in (\ref{lovelockmotion}) as 
$\delta^{bc_1\dots c_{2k}}_{ad_1\dots d_{2k}} = (1/(2k+1))(\delta_a^b\delta^{c_1\dots c_{2k}}_{d_1\dots d_{2k}}-2k\delta_{[d_1|}^b\delta^{c_1\dots c_{2k}}_{a|d_2\dots d_{2k}]})$
leads to the expression
$\mathcal{G}^{(k)}_a{}^b = \alpha_k(k\calrk_a{}^b-(1/2)\delta_a^b\calrk)$, where the Ricci-like tensor is given by
\begin{equation}\label{riccilike}
\calrk_a{}^b=\delta_{[d_1|}^b\delta_{a|d_2\dots d_{2k}]}^{c_1\dots c_{2k}} \, \,R_{c_1c_2}{}^{d_1d_2}\, \dots\,R_{c_{2k-1}c_{2k}}{}^{d_{2k-1}d_{2k}}
\end{equation}
For $k=1$ this reduces to the Ricci tensor, while taking the trace yields $\calrk_a{}^a = \calrk$.   

In order to obtain results for Lovelock gravity like those for Einstein gravity in $D=3$, we will also need a higher curvature Riemann-like tensor $\calrk_{ab}{}^{cd}$, such that 
$\calrk_a{}^b=\calrk_{ac}{}^{bc}$.
The similarity in structures between Lovelock and Einstein gravity, especially the Bianchi identity for the Einstein-like tensor\footnote{In the Einstein case, 
$\nabla_b G_a{}^b=0$ follows from contracting two pairs of indices in the underlying Bianchi identity for the Riemann tensor $\nabla_{[a}R_{bc]}{}^{cd}=0$.  It is natural to expect this to be the case at each higher curvature Lovelock order as well.}, strongly suggests that such a quantity should exist.  
However, it does not arise through consideration only of the Lovelock Lagrangian and equations of motion.

\section{Riemann-Lovelock  curvature}\label{5D}

We will focus first on the $k=2$ case.  In this case the curvature scalar and Ricci-like tensor in (\ref{lovelockterms}) and (\ref{riccilike}) are given by
\begin{align}\label{scalarlike}
\calr^{(2)} &= {1\over 6}\left( R_{ab}{}^{cd}R_{cd}{}^{ab}-4R_a{}^bR_b{}^a+R^2\right)\\
\label{ricciliketwo}
\calr_a^{(2)}{}^b 
&= {1\over 6}\left(R_{ae}{}^{cd}R_{cd}{}^{be} - 2R_{ad}{}^{bc}R_c{}^d - 2 R_c{}^bR_a{}^c +R_a{}^b R\right)
\end{align}
with the Einstein-like tensor given by $\calg^{(2)}_a{}^b = 12(\calr^{(2)}_a{}^b - {1\over 4}\delta_a^b\calr^{(2)})$.  
We are interested in simplifications of Lovelock  theory, including the $k=2$ term, that happen in $D=5$, the lowest dimension in which it makes a nontrivial contribution.  We expect that these simplifications will follow from an identity of the form (\ref{curv}), but involving quantities that are quadratic in the underlying curvature tensor.
In addition to the quantities (\ref{scalarlike}) and (\ref{ricciliketwo}), we expect this to also include a  tensor $\calr^{(2)}_{ab}{}^{cd}$ that has properties analogous to those of the  Riemann tensor.

For $k=1$, the Lovelock gravity quantities in section (\ref{lovelocksection}) reduce to the familiar quantities of Einstein gravity.  Generally speaking, to get from $k=1$ to $k=2$ in the formulas (\ref{lovelockterms}) and (\ref{lovelockmotion}), one adds another factor of the Riemann tensor and more indices on the antisymmetric delta-symbol.  In order to generalize the quantity $A_{ab}{}^{cd}$ defined in (\ref{basic}) to the $k=2$ case, we can try to do this same thing.
%
With a Lovelock perspective in mind, therefore, let us consider the quantity 
\begin{equation}\label{five}
\cala^{(2)}_{ab}{}^{cd}\equiv \delta_{abijkl}^{cdefgh}\,  R_{ef}{}^{ij}R_{gh}{}^{kl}.
\end{equation}
which
can be evaluated explicitly in any dimension and written in the form
\begin{equation}\label{lessbig}
\cala^{(2)}_{ab}{}^{cd} =  {1\over 15} \left( 6 \calr^{(2)}_{ab}{}^{cd} - 8 \delta^{[c}_{[a}\calr^{(2)}{}_{b]}{}^{d]} + \delta^{cd}_{ab} \, \calr^{(2)} \right) .
\end{equation}
Here the curvature scalar and Ricci-like tensor are the those given above in (\ref{scalarlike}) and (\ref{ricciliketwo}), while the new $4$-index tensor is given compactly in terms of antisymmetric delta symbols by
 $\calr^{(2)}_{ab}{}^{cd} =   \delta^{cd}_{[ij|}\,\delta^{efgh}_{ab|kl]} R_{ef}{}^{ij}R_{gh}{}^{kl}$  and more explicitly as
\begin{align}
\calr^{(2)}_{ab}{}^{cd}
=   & {1\over 18}\left(  RR_{ab}{}^{cd} + R_{ab}{}^{ef}R_{ef}{}^{cd} + 4 R_{[a}{}^{[c} R_{b]}{}^{d]}\label{riemannlike}
-4 R_{e[a}{}^{f[c}R_{b]f}{}^{d]e} \right .\\ &  \left .  + 4 R_{e[a}{}^{cd}R_{b]}{}^e + 4R_e{}^{[c}R_{ab}{}^{d]e}\right).\nonumber
\end{align}
One can verify that this tensor shares all the algebraic symmetries of the Riemann tensor, {\it i.e.} that 
$\calrr_{abcd}=-\calrr_{bacd}=-\calrr_{abdc}=+\calrr_{cdab}$ and $\calrr_{[abc]}{}^d=0$, and that contracting a pair of indices yields $\calrr_{ac}{}^{bc}=\calrr_a{}^b$.

At this point, we have already established our desired quadratic curvature identity in $D=5$, because in this dimension the quantity $\cala^{(2)}_{ab}{}^{cd}$ vanishes identically.  However, we will defer discussion of this until the next section.  We do this because the quantity $\calr^{(2)}_{ab}{}^{cd}$ turns out not to be the end of the story for $k=2$.  This is suggested by considering, in expectation that an analogue of the  differential Bianchi identity for the Riemann tensor should also hold, the quantity $\nabla_{[a}\calrr_{bc]}{}^{de}$.  This turns out to be non-vanishing and proportional to the divergence of a further tensor $\calrr_{abc}{}^{def}$, which again has algebraic symmetries like those of the Riemann tensor.

The full story can be obtained by considering the quantity 
\begin{equation}\label{big}
\calaa_{abcd}{}^{efgh} = \delta_{abcdmnop}^{efghijkl}\, R_{ij}{}^{mn}\, R_{kl}{}^{op}, 
\end{equation}
which can be evaluated in any dimension and written in the form
\begin{align}\label{bigg}
\calaa_{abcd}{}^{efgh} = {1\over 70}& \left\{ \delta_{abcd}^{efgh}\calrr - 16\delta_{[abd}^{[efg}\, \calrr_{d]}{}^{h]} + 36 \delta_{[ab}^{[ef}\, \calrr_{cd]}{}^{gh]}\right . \\  & \left .
 -16\delta_{[a}^{[e}\calrr_{bcd]}{}^{fgh]} +\calrr_{abcd}{}^{efgh}\right\}\nonumber
\end{align}
involving in addition to the tensors (\ref{scalarlike}), (\ref{ricciliketwo}) and (\ref{riemannlike}), the two further quantities
\begin{align}
\calrr_{abc}{}^{def} &= \delta_{[klm|}^{def}\, \delta_{abc|n]}^{ghij}\, R_{gh}{}^{kl}\, R_{ij}{}^{mn}
= {1\over 2}\left\{R_{[ab}{}^{[de}\, R_{c]}{}^{f]} - R_{n[a}{}^{[de}\, R_{bc]}{}^{f]n}\right\}\\
\calrr_{abcd}{}^{efgh} &= \delta_{mnop}^{efgh}\,\delta_{abcd}^{ijkl}\,  R_{ij}{}^{mn}\, R_{kl}{}^{op}
= R_{[ab}{}^{[ef}\, R_{cd]}{}^{gh]}.\label{antisym}
\end{align}
All the lower order tensors are obtained via contracting pairs of indices from the quantity $\calrr_{abcd}{}^{efgh}$, which we will call the $k=2$ Riemann-Lovelock tensor.  The Riemann-Lovelock tensor is clearly the maximal  such object that can be defined, since there are no more indices available on the two curvature tensors for anti-symmetrization.  

The properties of the Riemann-Lovelock tensor are now fully analogous to those of the Riemann tensor itself.
By construction the Riemann-Lovelock tensor has the symmetries $\calrr_{abcd}{}^{efgh}= \calrr_{[abcd]}{}^{efgh}=\calrr_{abcd}{}^{[efgh]}$.  It is straightforward to check that 
$\calrr_{abcdefgh}=\calrr_{efghabcd}$ and also the algebraic and differential Bianchi identities
\begin{align}
\calrr_{[abcde]}{}^{fgh} &=0\\
\nabla_{[a}\calrr_{bcde]}{}^{fghi}&=0\label{bianchi}
\end{align}
which follow from antisymmetry together with the identities satisfied by the Riemann tensor itself\footnote{The vanishing of $\nabla_{[a} \calrr_{bcde]}{}^{fghi}$ can also be obtained by taking considering 
$\nabla_f\cala_{abcde}^{(2)}{}^{fghij}=0$ with $\cala_{abcde}^{(2)}{}^{fghij}=\delta_{abcdeopqr}^{fghijklmn}\, R_{kl}{}^{op}\, R_{mn}{}^{qr}$.}.  The Bianchi identity $\nabla_b\calgk_a{}^b=0$ for the Einstein-like tensor then follows from contracting all but one pair of indices in (\ref{bianchi}).

Clearly, a similar construction may be carried out at each Lovelock order.  One finds that the $k$th order Riemann-Lovelock tensor given by
\begin{equation}\label{general}
\calrk_{a_1\dots a_{2k}}{}^{b_1\dots b_{2k}} = \delta_{a_1\dots a_{2k}}^{c_1\dots c_{2k}}\,\delta_{d_1\dots d_{2k}}^{b_1\dots b_{2k}}\, R_{c_1c_2}{}^{d_1d_2}\cdots
R_{c_{2k-1}{2k}}{}^{d_{2k-1}d_{2k}}
\end{equation}
shares the essential properties of the Riemann tensor.
It is manifestly totally antisymmetric in its up and down indices, and after lowering all indices is symmetric under interchange of the first and second sets of $2k$ indices.
Moreover, it can easily be shown to satisfy the Bianchi identities $\calrk_{[a_1\dots a_{2k}b_1]}{}^{b_2\dots b_{2k}}=0$ and 
$\nabla_{[a_1}\calrn_{a_2\dots a_{2k+1}]}{}^{b_1\dots b_{2k}}=0$.   We will see in the next section, that in sufficiently low dimensions the Riemann-Lovelock tensor may be expressed via a curvature identity in terms of some number of its contractions.

\section{Curvature identities}

At each Lovelock order $k$ there will be low dimensional curvature identities that determine the $k$th order Riemann-Lovelock tensor in terms of its contractions.  Let us start again by considering $k=2$.  The quantity $\calaa_{abcd}{}^{efgh}$ defined in (\ref{big}) vanishes identically for $D<8$.  From equation (\ref{bigg}) we see that this implies that the Riemann-Lovelock tensor $\calrr_{abcd}{}^{efgh} $ will be determined by some set of its contractions for all $D<8$.  Only for $D\ge8$  does it become fully general, in the same sense that the Riemann tensor itself does for $D\ge 4$.  In $D=4$, for example, the Riemann-Lovelock tensor $\calrr_{abcd}{}^{efgh} $ has only a single independent component and therefore must be determined fully in terms of $\calrr$.  In $D=5$, $\calrr_{abcd}{}^{efgh} $ has a symmetric tensor worth of independent components and is determined by $\calrr_a{}^b$, while in $D=6$ and $D=7$ it is determined in terms of its contractions $\calrr_{ab}{}^{cd}$ and $\calrr_{abc}{}^{def}$ respectively.  
The explicit forms for these relations can be determined by considering the quantities 
 \begin{equation}
 \calaa_{a_1\dots a_n}{}^{b_1\dots b_n}= \delta_{a_1\dots a_nghij}^{b_1\dots b_ncdef}\, R_{cd}{}^{gh}R_{ef}{}^{ij}
 \end{equation}
 for $n=1,2,3,4$, some, or all, of which vanish in dimensions $D=4,5,6,7$.
 
Let us work out the expression for the $k=2$ Riemann-Lovelock tensor in $D=5$ in detail.  From equation (\ref{lessbig}) we see that the identity $\calaa_{ab}{}^{cd}=0$ in $D=5$ yields  the relation
 \begin{equation}\label{again}
\calrr_{ab}{}^{cd}={1\over 6}\left( 8 \delta^{[c}_{[a}\calr^{(2)}{}_{b]}{}^{d]} - \delta^{cd}_{ab} \, \calr^{(2)}. \right)
\end{equation}
The relation $\calaa_{abc}{}^{def}=0$ can then be used in combination (\ref{again}) with to obtain an expression for $\calrr_{abc}{}^{def}$ in terms of $\calrr_a{}^b$ and $\calrr$, and finally using these in combination with $\calaa_{abcd}{}^{efgh}=0$ yields the expression 
\begin{equation}
\calrr_{abcd}{}^{efgh}= 16\delta_{[abc}^{[efg}\, \calrr_{d]}{}^{h]} - 3\delta_{abcd}^{efgh}\calrr .
\end{equation}
for the Riemann-Lovelock tensor in $D=5$ in terms of the Ricci-like tensor and its trace.  Identities for the $k=2$ Riemann-Lovelock tensor in the other dimensions $4\le D< 8$ can be similarly obtained.

The $k$th order Riemann-Lovelock tensor (\ref{general}), which has $2k$ indices anti-symmetrized over can be shown by similar means to be given in terms of its contractions in dimensions $2k\le D< 4k$.  In $D=2k$, the Riemann-Lovelock tensor has a single independent component and is determined by the curvature scalar $\calrk$, while in $D=2k+1$ it is determined by the Ricci-like tensor $\calrk_a{}^b$. 

\section{Lovelock gravity in low dimensions}\label{solutions}

Amongst the curvature identities discussed in the last section, we have seen that the $k$th Riemann-Lovelock tensor 
$\calrk_{a_1\dots a_{2k}}{}^{b_1\dots b_{2k}}$ is determined by the Ricci-like tensor $\calrk_a{}^b$ in dimension $D=2k+1$.  
Pure $k$th order Lovelock theory in $D=2k+1$, in which $\callk$ is the sole term in the Lovelock Lagrangian, provides the most direct parallel to Einstein gravity in $D=3$.  For the pure theory, the field equation is simply $\calrk_a{}^b=0$ and it follows that the $k$th order Riemann-Lovelock tensor must vanish on all solutions in this theory.  We will call such spacetimes $k$th order Riemann-Lovelock flat.

The solutions to pure Einstein gravity ({\it i.e.} with vanishing cosmological constant) in $D=3$ are flat.  This implies that the static, spherically symmetric solutions are the conical space times \cite{Deser:1983tn}
\begin{equation} 
ds_3^2 = -dt^2 +dr^2 +\alpha^2r^2 d\theta^2
\end{equation}
which for $\theta\equiv\theta+2\pi$ have deficit (excess) angle for $\alpha^2<1$ ($\alpha^2>1$).  For $\alpha^2\neq 1$, these spacetimes are locally flat away from the origin, where there is a conical singularity.  

The static, spherically symmetric solutions of pure $k$th order Lovelock gravities  can be found in reference \cite{Crisostomo:2000bb}.  In $D=2k+1$ these are spacetimes of missing (or excess) solid angle, given by
\begin{equation}\label{missingangle}
ds_{2k+1}^2 = -dt^2 +dr^2 + \alpha^2 r^2 d\Omega_{2k-1}^2
\end{equation}
where $d\Omega_N^2$ is the metric on the round, unit $N$-sphere.  We can check that these spacetimes are indeed $k$th order Riemann-Lovelock flat. 
Denoting the coordinates on the sphere by $x^i$, the only non-vanishing components of the Riemann tensor for (\ref{missingangle}) have all indices tangent to the sphere and are given by
\begin{equation}
R_{ij}{}^{kl} = {2\over \alpha^2\, r^2}(1-\alpha^2)\delta_{ij}^{kl}
\end{equation}
Since the sphere has dimension $2k-1$ and the formula for $\calrk_{a_1\dots a_{2k}}{}^{b_1\dots b_{2k}}$ requires antisymmetrization over $2k$ indices, it follows that  
$\calrk_{a_1\dots a_{2k}}{}^{b_1\dots b_{2k}}$ vanishes identically and that the missing (excess) solid angle spacetimes (\ref{missingangle}) are indeed $k$th order Riemann-Lovelock flat.  It is an intriguing that, despite their non-trivial curvature, these simple generalizations of conical space times are nonetheless `flat' in this sense.

\section{Conclusions and further questions}

In this paper, we have introduced a notion of Riemann-Lovelock curvature, characterized by the Riemann-Lovelock tensors (\ref{general}) and examined some of its consequences in low dimensions.  A number of further questions come to mind as well, which can serve as the basis for future investigations.  The most important of these is to ask whether the curvature identities found here have implications for the solutions of general Lovelock theories, beyond the class of pure theories considered in the preceding section.  In the case of $D=3$, the identity (\ref{curv}) also implies that solutions to Einstein gravity with a non-vanishing cosmological constant necessarily have constant curvature \cite{Deser:1983nh}.  Since this represents the most general Lovelock theory in $D=3$, one might expect strong results on the curvatures of solutions to general Lovelock gravities to exist in all odd dimensions.  

Preliminary results indicate that this is not the case.  However, Einstein gravity in $D=3$ can also be thought of as an example of a special class of Lovelock theories known as Chern-Simons theories (see {\it e.g.} reference \cite{Crisostomo:2000bb} for a discussion), which have a unique constant curvature vacuum.  Preliminary results do point to a generalization of the results of section (\ref{solutions}) applying in the Chern-Simons case.
Vacuum Einstein gravity in $D=3$ and pure $k$th order Lovelock theory in $D=2k+1$ are limiting cases of Chern-Simons gravity in which the vacuum curvature is taken to zero. 

Some further possible questions are the following.
It is well known that in $D=2$ coordinates can always be found such that locally the metric is conformal to a constant curvature one.   Is there some analogue of this result in $D=2k$, {\it e.g.} can the metric always be put in a form that is conformal to one that has constant $k$th order Riemann-Lovelock curvature?  

It is also well known that the Weyl tensor vanishes in $D=3$ by virtue of the identity (\ref{curv}).  A traceless, Weyl-like tensor $\calck_{a_1\dots a_{2n}}{}^{b_1\dots b_{2n}}$ may presumably be constructed from the $k$th order Riemann-Lovelock tensor and its traces.  This tensor also presumably vanishes in $D=2k+1$ and transforms simply under conformal transformations.  It would be interesting to see what its implications are.  For example, the square of the Weyl tensor is added to the gravitational Lagrangian to obtain the possibly renormalizable `critical gravity' theory in $D=4$ \cite{Lu:2011zk}.  This work has also been related \cite{Lu:2011ks} to the attempt to derive Einstein gravity from conformal gravity via imposition of boundary conditions in \cite{Maldacena:2011mk}.  It seems plausible that the $k$th order Weyl-like tensor can be used for similar purposes in the Lovelock gravity context.

Finally, it would be desirable to more fully address the original question of the similarity of Einstein gravity in $D=3$ to Lovelock gravity in $D=2k+1$.  We have discussed the analogous curvature identities that hold and some of their implications for  solutions.  However, we have not addressed the issue of local degrees of freedom.  In the case of $D=3$ Einstein gravity there are none.  Spacetime is flat.  However, the condition of Riemann-Lovelock flatness that holds for solutions of pure $k$th order Lovelock gravity in $D=2k+1$ allows for non-trivial Riemannian curvature.  It would be interesting to understand how the condition impacts gravitational radiation in these theories.

\textbf{Acknowledgements:}  The author would like to thank Xian Camanho, Christos Charmousis, Lorenzo Sorbo, Jennie Traschen and Jorge Zanelli for helpful conversations and the organizers of the \textit{Gravity - New perspectives from strings and higher dimensions} workshop at the Centro de Ciencias de Benasque Pedro Pascual for hospitality during part of this project.

\textbf{Note Added:}
After this work was largely complete, we found an earlier reference \cite{Dadhich:2008df}, which includes a substantial part of the story of Riemann-Lovelock curvature story presented here.
In our notation, this work deals with the quantity $\calrk_{ab}{}^{cd}$, which in terms of number of indices is the most direct higher curvature analogue of the Riemann tensor.  We have additionally seen here that this quantity naturally arises via contraction from the Riemann-Lovelock tensor (\ref{general}) which, in addition to sharing the algebraic symmetries of the Riemann tensor, also satisfies the Bianchi identity $\nabla_{[a_1}\calrn_{a_2\dots a_{2k+1}]}{}^{b_1\dots b_{2k}}=0$.  

Very recently, as this drafting was being finished, reference \cite{Dadhich:2012cv} has appeared as well, which asks the same central question as ours, regarding an analogue of the behavior of Einstein gravity in $D=3$ for Lovelock gravity in $D=2k+1$, with similar results to ours such as those on missing solid angle space times.  The results of \cite{Dadhich:2012cv} are again based on the contracted quantity $\calrk_{ab}{}^{cd}$ and are therefore somewhat less comprehensive\footnote{Although reference \cite{Dadhich:2012cv} also contains additional results and discussion of the Lovelock case with inclusion of  a non-zero cosmological constant.}.

\end{document}